\documentclass[pdflatex,sn-mathphys-num]{sn-jnl}

\usepackage[framemethod=TikZ]{mdframed}
\usepackage{courier} 
\usepackage{float}
\usepackage{tcolorbox}
\usepackage{graphicx}
\usepackage[english]{babel} 
\usepackage{float}
\usepackage{url}
\usepackage{blindtext}
\usepackage{threeparttable}
\usepackage{booktabs}
\usepackage{graphicx}%
\usepackage{multirow}%
\usepackage{amsmath,amssymb,amsfonts}%
\usepackage{amsthm}%
\usepackage{mathrsfs}%
\usepackage[title]{appendix}%
\usepackage{textcomp}%
\usepackage{manyfoot}%
\usepackage{booktabs}%
\usepackage{algorithm}%
\usepackage{algorithmicx}%
\usepackage{algpseudocode}%
\usepackage{listings}%
\newtcolorbox{codebox}[1][]{
    colback=codegrey,
    colframe=black,
    fonttitle=\bfseries,
    title=Code Snippet,
    #1
}

\usepackage[normalem]{ulem}

\begin{document}

\title{Watts per event: evaluating Sustainability of HEP Event Generators beyond the LHC era}

\author[1]{\fnm{Szabolcs} \sur{Molnár}}\email{molnar.szabolcs@wigner.hun-ren.hu}

\author[1]{\fnm{Gábor} \sur{Bíró}}\email{biro.gabor@wigner.hun-ren.hu}

\author[2]{\fnm{Gábor} \sur{Papp}}\email{gabor.papp@ttk.elte.hu}

\author[1]{\fnm{Gergely Gábor} \sur{Barnaföldi}}\email{barnafoldi.gergely@wigner.hun-ren.hu}

\affil[1]{\orgdiv{Department of Theoretical Physics}, \orgname{HUN-REN Wigner Research Centre for Physics}, \orgaddress{\street{29-33 Konkoly-Thege Miklós rd.}, \city{Budapest}, \postcode{1121}, \country{Hungary}}}

\affil[2]{\orgdiv{Department of Theoretical Physics}, \orgname{Eötvös Loránd University}, \orgaddress{\street{Pázmány Péter Sétány 1/A}, \city{Budapest}, \postcode{H-1117}, \country{Hungary}}}

\abstract{
The development, tuning and operation of Monte Carlo event generators beyond the LHC era require vast amount of resources. In this study we investigate the sustainability of these software with a containerized set of tools (named \texttt{77rev/propripy}), by benchmarking the \texttt{HIJING++} heavy-ion Monte Carlo event generator. We analyze the performance of various CPU architectures and show that by choosing the level of multithreading properly, the cost of event generation can be optimized. The presented approach can reduce the energy footprint of high-energy physics event generators and therefore alleviate the ever-increasing, ubiquitous computational challenges.
}

\maketitle

\section{Introduction}
\label{sec:intro}

The upcoming High-Luminosity LHC (HL-LHC), and later the Future Circular Collider (FCC) era present unprecedented computational challenges, as the required scale of simulated data is expected to increase by an order of magnitude~\cite{ZurbanoFernandez:2020cco, LHeC:2020van, FCC:2018evy, FCC:2018byv}. Monte Carlo (MC) event generation remains one of the most resource-intensive components of the HEP computing pipeline, often consuming a majority of the total CPU power provided by the Worldwide LHC Computing Grid (WLCG)~\cite{Bird:2005js}. As the community shifts toward a "carbon-aware" computing model, the metric for success is evolving from simple event throughput to sustainable efficiency~\cite{ATLAS:2025sgg}. This becomes especially important if the globally increasing hardware prices are showing an increasing trend.

Monte Carlo event generators are computational tools that simulate the complete evolution of particle collisions. The calculations are separated into two parts: hard and soft processes. Quantum Chromodynamics (QCD) is a non-abelian gauge theory that describes the strong interaction~\cite{ALICE:2022wpn}. On high energies QCD is perturbative, while on low energies where the coupling becomes strong it is non-perturbative. This non-perturbative energy domain is what we refer to as soft QCD and is handled by effective theories or phenomenology. Overall, these properties make it challenging to model particle collisions accurately. The non-perturbative nature of the soft processes and the requirement to model both hard and soft parts necessitate many parameters, some of which are non-physical. For the model to have predictive power, all of the parameters have to be tuned~\cite{Buckley:2009bj, Lazzarin:2020uvv} using experimental data, e.g. from the LHC~\cite{Evans:2008zzb}.

In the following sections, we introduce a scoring system to measure efficiency, and a specifically developed toolbox for its evaluation. We illustrate the methodology by presenting the results measured via a heavy-ion Monte Carlo event generator at LHC and FCC energies.

\section{Efficiency Scoring}
\label{sec:resource}

Current HEP computing demands a departure from traditional "time-to-completion" metrics. As software grows more complex---incorporating deep multithreading and vectorization---the hardware utilization profile changes significantly. Modern workloads require a benchmarking approach that prioritizes real-world application performance over synthetic instruction sets. By focusing on the operational efficiency of specific production tasks, researchers can better understand how hardware limitations, such as memory bandwidth and thermal throttling, impact the overall throughput of large-scale simulation campaigns.

To evaluate the sustainability of an event generator, a "price" metric can be defined to measure the total energy consumed to produce a single physics event. The HEPScore23 (HS23) framework shifts benchmarking toward throughput-per-watt metrics using real production workloads~\cite{Szczepanek:2024uun}. This energy cost can be achieved by correlating the average consumed power (P) in Watts, with the event throughput ($T_{\mathrm{event}}$) in events per second:

\begin{equation}
    E_{\mathrm{event}} = \frac{P_{\mathrm{avg}}}{T_{\mathrm{event}}}.
    \label{eq:cost}
\end{equation}

This allows us to keep track of the efficiency of any event generator across different levels of parallelism, assuming that for the benchmarking the CPUs were dedicated to this task only. In turn, we can ascertain which level of parallelism (i.e., how many threads) is optimal.

\section{Motivation: Monte Carlo tuning}
\label{sec:motivation}

Simulating heavy ion collisions---and consequently the tuning of a heavy-ion event generator---is especially challenging. 
Parameter tuning is an iterative process, where one minimizes the cost function. For this step we are using the \texttt{Professor} tool~\cite{Buckley:2009bj}. The process begins with a stochastic sampling from a defined range of parameter values, resulting in a large set of parametrisations that the event generator has to be evaluated on.
This results in a vast amount of generated data that needs to be used for the MC parameter optimisation.

This process involves many tools that need to interact with each other, e.g.,
in our example the Monte-Carlo step produces files in \texttt{HepMC3} format, which is processed via a \texttt{Rivet}~\cite{Bierlich:2024vqo} analysis, followed by creation a physically relevant data structures through \texttt{YODA} files.
In the last step, \texttt{Professor} can be invoked to interpolate the results and perform the cost function minimization---quite often in several passes, with variable weights, limit ranges of parameters and initial values.

\section{Methodology}
\label{sec:method}

As the whole tuning process involves tedious and technically complex processes to be performed many times, a specific toolbox has been built in order to facilitate the tuning.
The toolbox is available as a \texttt{Docker} image  (on Docker Hub), named \texttt{77rev/proripy}~\cite{docker_proripy} (\textbf{Pro}fessor, \textbf{Ri}vet and \textbf{Py}thia8). The image makes it easy to use the included tools without step-by-step installation and maintenance. Furthermore, the image includes several compilers, making it possible to debug and run code inside the container.
Additionally, this tool can be used to determine the optimal parallelisation level of the whole framework on a given hardware, which results in a more efficient procedure. The current versions of tools in the image are summarized in Table~\ref{tab:tools}.

The default tune of \texttt{Pythia8}--the Monash tune~\cite{monash}--requires over 30 pages just to present the results. It has different sets of parameters for different goals and fine-tunings, tuned for a lot more results than what is presented in this work. Presenting a full tuning process is outside the scope of this paper.

\begin{table}[h]
\caption{Included main packages and their versions}\label{tab:tools}
\begin{tabular*}{\textwidth}{@{\extracolsep\fill}lccccccc}
\toprule
Package & Rivet & Pythia8 & Professor & ROOT & YODA & HepMC3 & GDB \\
\midrule
Version & 4.1.0 & 8.316 & 2.4.2 & 6.36.04 & 2.1.0 & 3.3.1 & 12.1 \\
\midrule
Package & FastJet5 & HighFive & Ruby & C/C++ & Rust & Python & Go \\
\midrule
Version & 3.4.3 & 2.10.1 & 3.0.2p107 & 11.4.0 & 1.28.2 (rustup) & 3.10.12 & 1.18.1 \\
\botrule
\end{tabular*}
\end{table}

In this study, we demonstrate the importance of optimal parallelisation level, motivated by the high computation requirements of the tuning process, by benchmarking and evaluating several hardware architectures using the \texttt{77rev/proripy} toolbox. Additionally, preliminary tuning results of the \texttt{HIJING++} (Heavy Ion Jet INteraction Generator) MC event generator, developed in C++ with CPU multithreading capabilities, are shown as an illustration of the method~\cite{Albacete:2017qng, Barnafoldi:2017jiz, Papp:2018qrc, Biro:2018ntr, Biro:2019ngd, Biro:2019ijx, Sjostrand:1982fn, Sjostrand:2014zea, Skands:2014pea, Christiansen:2015yqa, Sjostrand:2017cdm, Bierlich:2018xfw, Bierlich:2022pfr}. 

We tested the performance generating 500,000 pp events and 10,000 Pb-Pb events with detailed \texttt{HepMC3} formatted output, while no other processes (besides the system-processes running in the background) were running to ensure the accurate power measurements. 

The script for benchmarking is included in the image, 
located at \texttt{/home/devuser/Scripts}. It also requires the \texttt{power.sh} script,
which has to be in the same folder. The script has to be run via a privileged docker container with root permissions 
so that the \texttt{powerstat} linux package can have access to the hardware. The way the benchmarking scripts work, along with other technical details is explained on the Docker image's Gitlab page~\cite{toolboxgitlab}.

\subsection{Investigated architectures}
\label{sec:perf}

Here we present the impact of CPU multithreading on performance, as well as show the power consumption of multiple CPU architectures. With this, we can answer the question of which CPU is the most efficient (power consumption vs performance)---and, more importantly, what is the optimal level of parallelisation on a given architecture. The CPUs tested (single- and multithread) are shown in Table~\ref{tab:cpu_specs}. By power consumption, we mean the CPU's momentary power consumption specifically (not the Thermal Design Power - TDP or Package Power Tracking - PPT).  

TDP is a thermal engineering specification that tells cooler manufacturers how much heat the processor is expected to generate under sustained workloads so they can design adequate cooling. In contrast, PPT is the actual electrical power cap enforced by the CPU’s own management firmware. On AMD platforms, PPT is a hard average wattage limit for the entire socket (cores, I/O die, memory controller), typically set at about $1.35$ times the TDP, and the processor will not exceed it (for more than a very short burst at most). The \texttt{powerstat} (v$0.02.27$) tool measures the real‑time power consumption of the CPU package by reading the RAPL (Running Average Power Limit) hardware energy counters---the very same model‑specific registers the processor uses to enforce PPT---and it reports the actual power (watts) being drawn at that moment. Consequently, \texttt{powerstat} shows the power that is bounded by (but is not the value of) the PPT limit itself.

The drives' and the RAM's power requirement would slightly elevate the values. However, the majority of the power consumption comes from the CPU anyway and to keep it from being overly complicated, it is best if we neglect the rest of the components (furthermore, it would also require different tools to measure the power consumption of other components). 

\begin{table}[h]
\caption{CPU Details and specifications}
\label{tab:cpu_specs}
\begin{tabular*}{\textwidth}{@{\extracolsep{\fill}}l c r r r c}
\toprule
\textbf{CPU} & \textbf{Cores/Thr.} & \textbf{Clock (Base/Boost)} & \textbf{Cache} & \textbf{TDP} & \textbf{Year} \\
\midrule
Intel Xeon E5-2650~\cite{xeon}   & 8C/16T  & 2.00 / 2.80 GHz & 20 MB  & 95 W  & 2012 \\
AMD EPYC 7502P~\cite{7502P}      & 32C/64T & 2.50 / 3.35 GHz & 144 MB & 180 W & 2019 \\
AMD Ryzen 7 8845HS~\cite{8845HS} & 8C/16T  & 3.80 / 5.14 GHz & 24 MB  & 54 W  & 2023 \\
AMD EPYC 4585PX~\cite{4585PX}    & 16C/32T & 4.30 / 5.75 GHz & 144 MB & 170 W & 2025 \\
\bottomrule
\end{tabular*}
\end{table}

It was ensured that the memory (RAM) is not a bottleneck in either cases. When running this event generator it is important to have at least 32GB if not more, otherwise the RAM becomes a hard bottleneck (especially if it is DDR4 or older). The Intel Xeon CPU was paired with 32GB DDR3 RAM which was more than sufficient for the runs.

The frequency was governed by the kernel's built in \texttt{CPUFreq} subsystem. It has 3 parts, the core, scaling governors and scaling drivers. Scaling governors use algorithms to estimate the required CPU capacity so the subsystem can adjust it accordingly.

\section{Results}
\label{sec:results}

The tuning of Monte Carlo generators---the process of adjusting physical parameters to match experimental data---is an iterative and computationally expensive task requiring billions of events to be generated. If the underlying generation configuration is not optimized for energy efficiency, the cumulative carbon cost of a tuning campaign can be staggering. By integrating power-aware metrics into the tuning workflow (similarly to the HS23 metric), one can make informed decisions about the "price" of precision, aligning with the long-term computational strategy of the High Luminosity LHC era. 
Below we show that by the configuration that yields the highest efficiency score, effectively treating energy as a finite budget, a more optimal tuning strategy can be utilized. The benchmarks were run for 16 threads but we only show 10 on the compact figures for better visibility.

\subsection{HL-LHC era}
\label{sec:results1}

\begin{figure}[h]
  \centering
  \includegraphics[width=1\linewidth]{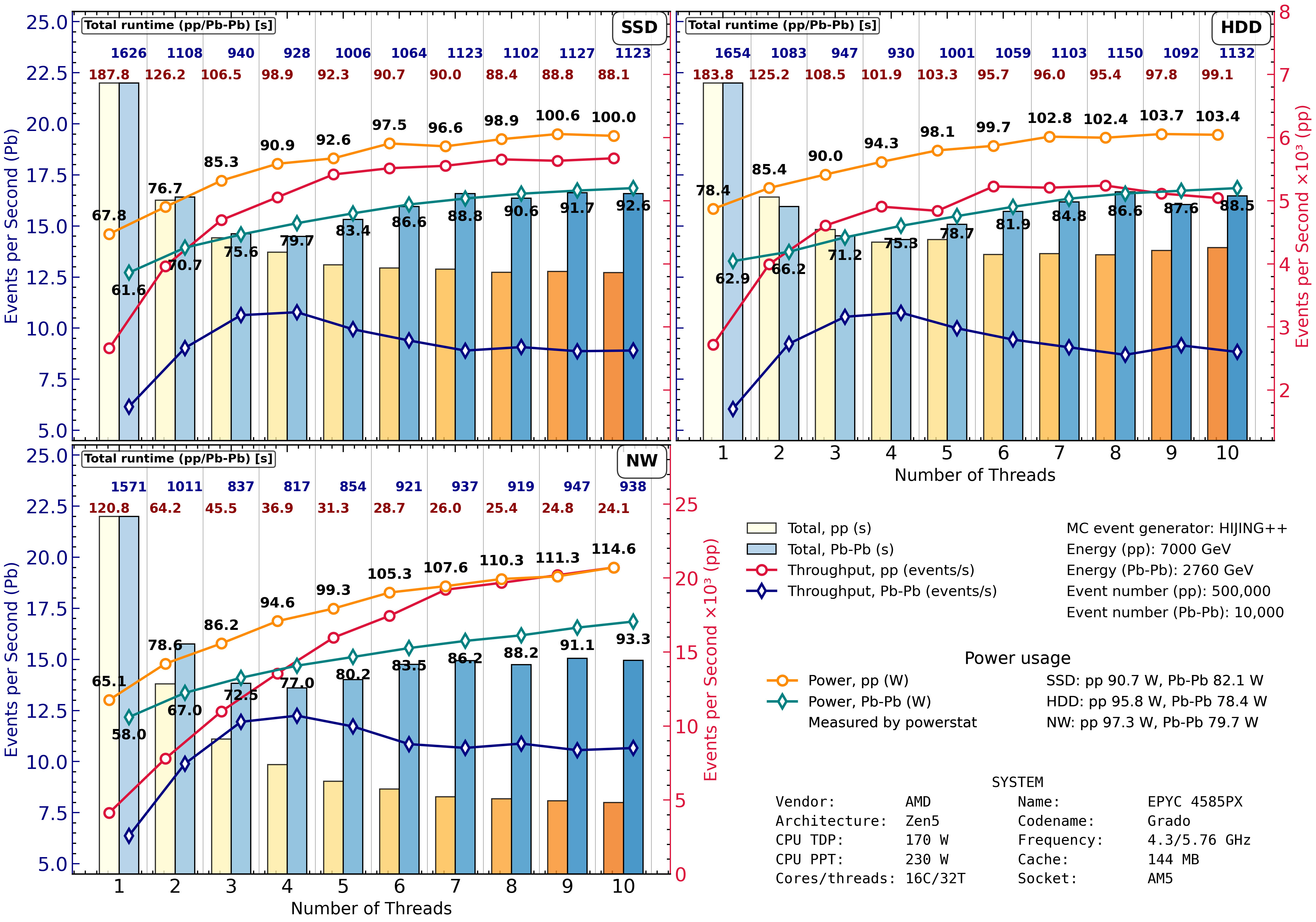}
  \caption{Performance metrics of the AMD EPYC 4585PX at 7 TeV pp and 2.76 TeV Pb-Pb collisions.}
  \label{fig:4585px_main}
\end{figure}

On Fig.~\ref{fig:4585px_main} we show the summarized, compact results for AMD EPYC 4585PX on HL-LHC energies. 
One might have expected that the drive type will play a significant role in the process (be a limiting factor in the case of HDDs), but the findings don't necessarily support this claim. The HDD did indeed decrease the speed (by around $500$ events/sec) but by far less than we expected. The no-write (NW) case simply shows the raw capability of the CPU without any limiting factor. Thus we can conclude that the limiting factor (in the case of newer CPUs) is the bandwidth between the components (CPU-memory and so on), which is not something the user has control over.

Comparing the results of the Ryzen 7 8845HS and the EPYC 4585PX against the older CPUs might falsely lead us to believe that the difference in clockspeed is the most important metric. The deeper reason for this difference is the CPU architecture and how efficiently can the CPU be used by the program we are running (the newer CPU the better). These comparisons will be evaluated in detail in Section \ref{sec:comp} and in Appendix~\ref{sec:app-plottab}.

The trends we see on the plot (power consumption and speed) is also heavily influenced by the MC event generator itself, how it is optimized, what CPU instructions could be utilized fully and so on. 

\begin{figure}[h]
  \centering
  \includegraphics[width=1.0\linewidth]{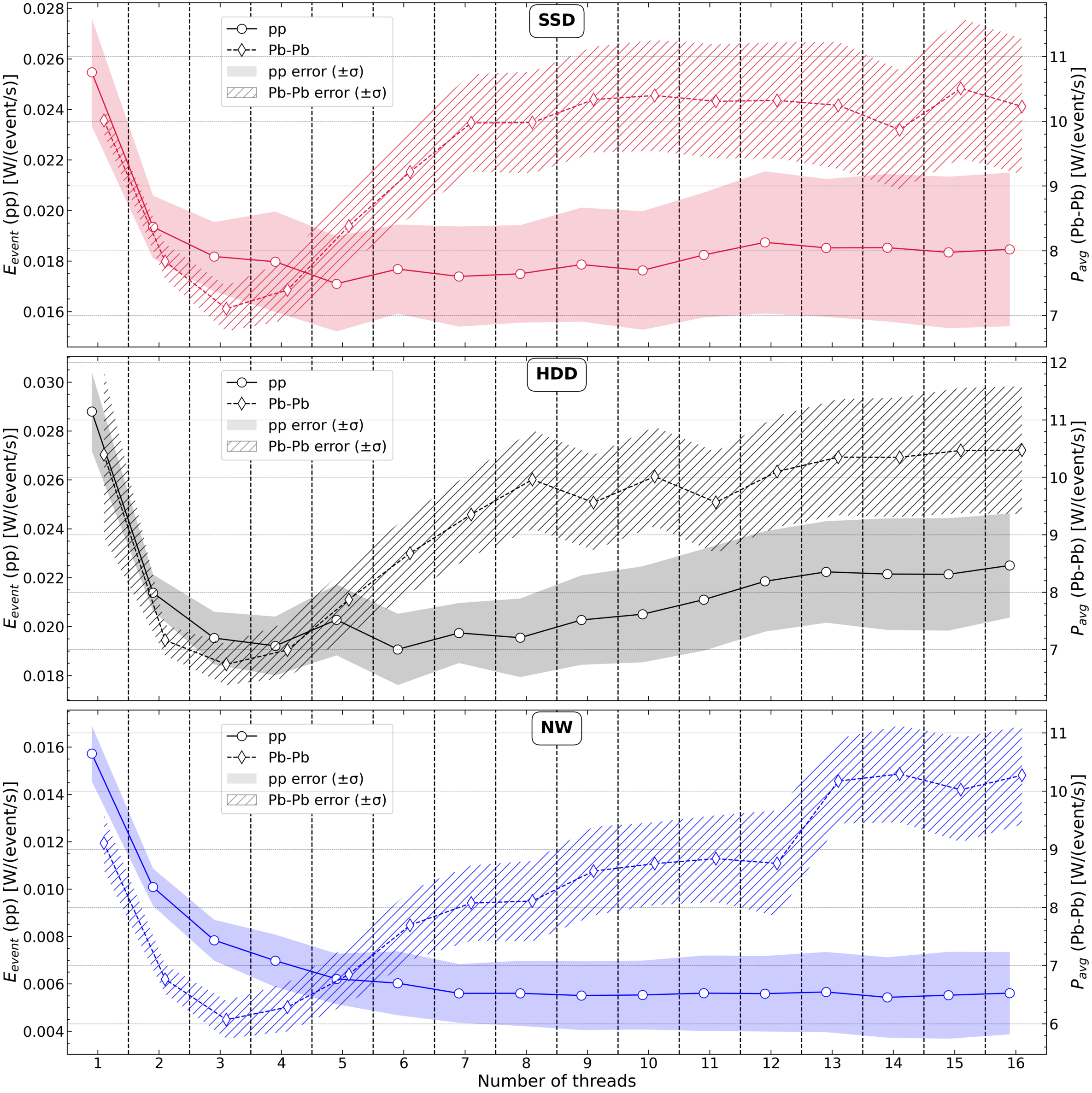}
  \caption{Cost per event plot of the AMD EPYC 4585PX in all 3 cases - HL-HLC}
  \label{fig:effy}
\end{figure}

On Fig.~\ref{fig:effy} the power cost of the event generation is plotted as defined in Eq. (\ref{eq:cost}), with respect to the number of used threads. In every investigated case there is a clear optimum, a most advantageous thread number to choose per drive type, with the best event/s to power ratio. The results of all investigated CPU architectures, along with the  changes of the runtimes of the optimal configurations compared to the fastest ones, are also summarized in Table~\ref{tab:cpu_res}.

\begin{table}[h]
\caption{Results table showing which number of threads is the fastest and most optimal for all tested CPUs. The measured times for the fastest values were taken as the reference (100\%).}
\label{tab:cpu_res}
\begin{tabular*}{\textwidth}{@{\extracolsep{\fill}}l c c c c r r}
\toprule
\multirow{2}{*}{\textbf{CPU}} &
  \multicolumn{2}{c}{\textbf{Fastest}} &
  \multicolumn{2}{c}{\textbf{Most optimal}} &
  \multicolumn{2}{c}{\textbf{Change in time}} \\
& pp & Pb-Pb & pp & Pb-Pb & pp & Pb-Pb \\
\midrule
Intel Xeon E5-2650 & 2 & 3 & 2 & 3 & 100.0\% & 100.0\% \\
AMD EPYC 7502P     & 2 & 4 & 2 & 4 & 100.0\% & 100.0\% \\
AMD Ryzen 7 8845HS & 6 & 5 & 4 & 3 & 104.1\% & 113.9\% \\
AMD EPYC 4585PX - SSD & 15 & 4 & 5  & 3 & 104.9\% & 101.3\%\\
AMD EPYC 4585PX - HDD & 8  & 3 & 6  & 3 & 113.7\% & 100.0\%\\
AMD EPYC 4585PX - NW  & 10 & 4 & 10 & 4 & 100.0\% & 100.0\%\\
\bottomrule
\end{tabular*}
\end{table}

We can observe a non-linear trend in the decrease of cost. The thread handling and additional steps that are required for multithreading, as apparent, consume time and/or energy in a rate which cancels out or even overtakes the gain from the multithreading itself.

Interestingly, in Table \ref{tab:cpu_res} one can see that the difference between the fastest and optimal configurations becomes significant only for the newer CPU types.
The runtime difference in all cases is almost negligible.
In case of the newest EPYC 4585PX CPU, these differences are more pronounced: for pp collisions, the number increase of threads becomes sub-optimal above $N_{threads}>5$ for the SSD setup, while the fastest configuration is reached at $N_{threads}=15$. This indicates that a significant amount of energy can be saved by optimizing not to the runtime, but to the energy efficiency. 

Additionally, for Pb-Pb collisions at the same CPU type,  the most optimal setup is different only in the case of the SSD.

In Section \ref{sec:comp} the runtime speed of the optimal and fastest configurations will be compared in detail.

\subsection{FCC era}
\label{sec:results2}

\begin{figure}[h]
  \centering
  \includegraphics[width=1\linewidth]{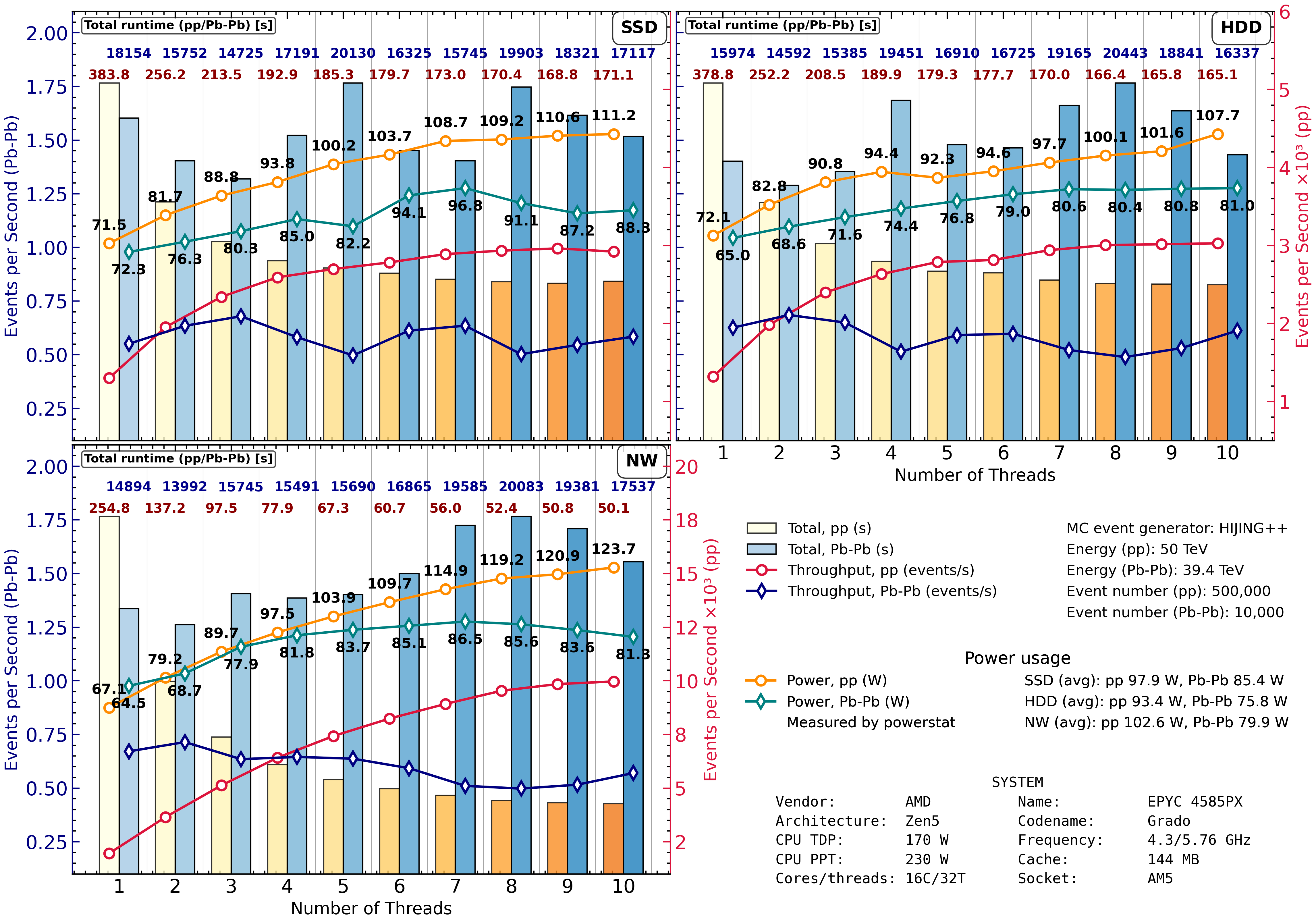}
  \caption{Performance metrics of the AMD EPYC 4585PX at 50 TeV pp and 39.4 TeV Pb-Pb collisions}
  \label{fig:4585px_FCC}
\end{figure}

Generally, higher energies, especially in the case of heavy-ion collisions, require far more computational power. Overall, however, we can see on Fig.~\ref{fig:4585px_FCC} that the same trends are preserved as in Fig.~\ref{fig:4585px_main}.

\begin{figure}[h]
  \centering
  \includegraphics[width=1.0\linewidth]{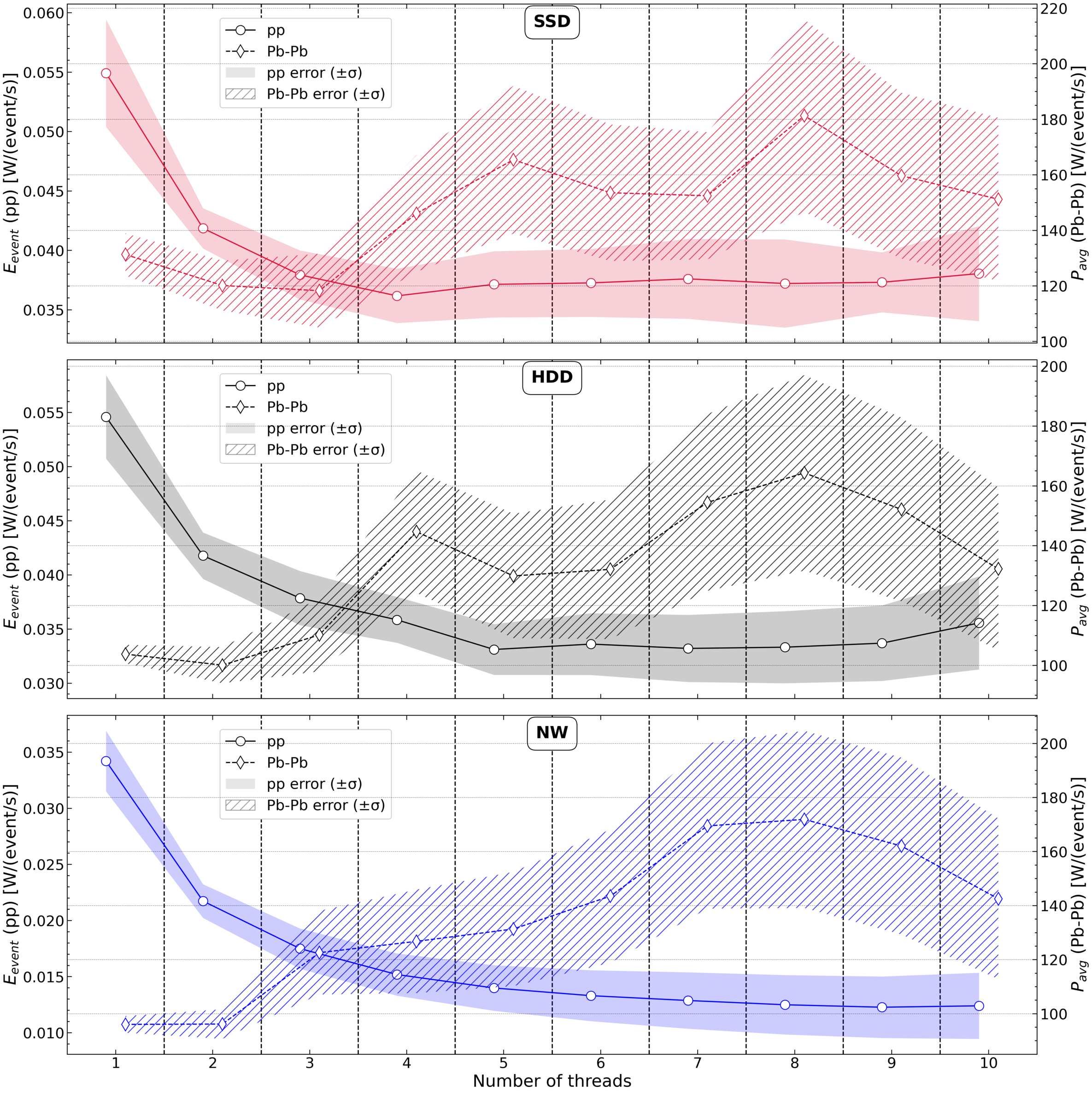}
  \caption{Cost per event plot of the AMD EPYC 4585PX, FCC energies, up to 10 threads.}
  \label{fig:effy_FCC}
\end{figure}

Figure~\ref{fig:effy_FCC} shows the cost of each event with respect to the used number of threads, for SSD, HDD and NW cases respectively.

It is immediately visible that these collision energy regimes are much more challenging to compute and the optimal level of multithreading is different than the LHC energies---moreover, in the case of Pb-Pb collisions the most optimal thread count was already found right away. These results can be seen in Table~\ref{tab:4585px_opt_FCC}.

\begin{table}[h]
\caption{Results table showing which number of threads is the fastest and most optimal for the AMD EPYC 4585PX on FCC energies. The measured times for the fastest values were taken as the reference (100\%).}
\label{tab:4585px_opt_FCC}
\begin{tabular*}{\textwidth}{@{\extracolsep{\fill}}l c c c c r r}
\toprule
\multirow{2}{*}{\textbf{CPU}} &
  \multicolumn{2}{c}{\textbf{Fastest}} &
  \multicolumn{2}{c}{\textbf{Most optimal}} &
  \multicolumn{2}{c}{\textbf{Change in time}} \\
& pp & Pb-Pb & pp & Pb-Pb & pp & Pb-Pb \\
\midrule
AMD EPYC 4585PX - SSD    & 9 & 3 & 4 & 3 & 114.3\% & 100\%\\
AMD EPYC 4585PX - HDD    & 10 & 2 & 5 & 2 & 108.6\% & 100\%\\
AMD EPYC 4585PX - NW     & 10 & 2 & 10 & 2 & 100\% & 100\%\\
\bottomrule
\end{tabular*}
\end{table}

\section{Tuning results}
\label{sec:tuningresults}

In the following subsections we present the pp tuning results for LHC energies, and we show the preliminary predictions of \texttt{HIJING++} for 39.4 TeV Pb-Pb collisions.

\subsection{HL-LHC era}
\label{sec:tune1}

\begin{figure}[h]
  \centering
  \includegraphics[width=0.9\linewidth]{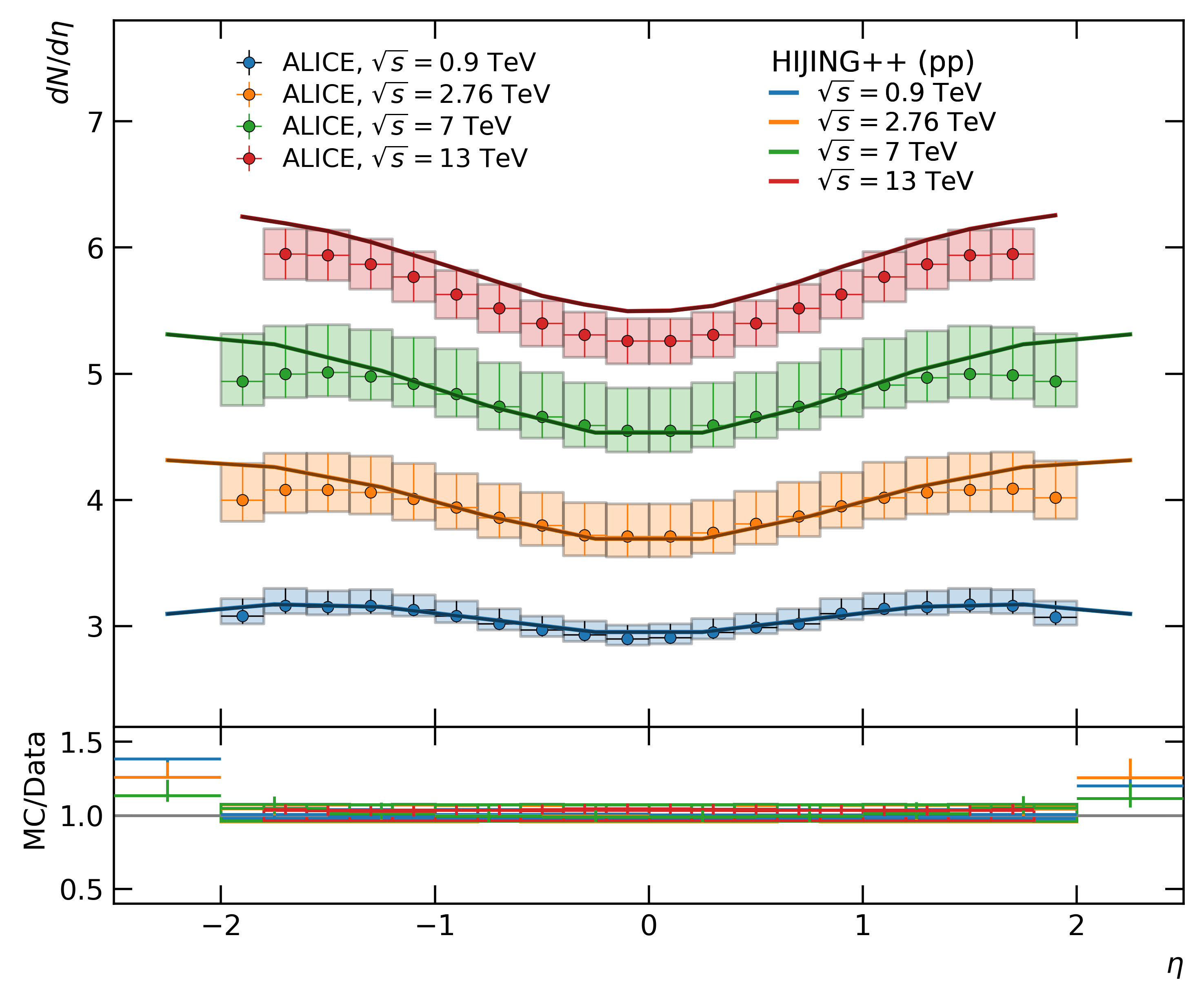}
  \caption{Pseudorapidity distributions on HL-LHC energies~\cite{PhysRevC.83.024913, ALICE:2015olq, ALICE:2018vuu, ALICE:2015qqj}}
  \label{fig:tuneeta}
\end{figure}

This section contains some recent results of the parameter tuning of \texttt{HIJING++}, for which the toolbox docker image was utilized. The results for pseudorapidity distribution are presented in Fig.~\ref{fig:tuneeta}. As it was detailed in Sec.~\ref{sec:method}, this process was done via the tuning tool \texttt{Professor}, at $\sqrt{s}=0.9$ TeV, $2.76$ TeV, $7$ TeV and $13$ TeV collision energies in proton-proton systems~ \cite{PhysRevC.83.024913, ALICE:2015olq, ALICE:2018vuu, ALICE:2015qqj}. Overall 6 parameters were tuned during this process, with $\mathcal{O}(1000)$ sampled parametrization configs, 1-3 iterations, with 5 million events at each config---this results in $\mathcal{O}(6\cdot 10^{10})$ events for this simple tuning campaign. This is a very minimal setup, done for demonstrative purposes. A full, more rigorous tuning process would take far more time, and would require significantly more events.

The tuning process was focused only on the pseudorapidity distributions at multiple energies, see Fig.~\ref{fig:tuneeta}. 
We can state, that the technical side of the process was made very smooth by the Toolbox. We also managed to estimate the power requirement of the illustrated tuning campaign, 
which is presented in Table~\ref{tab:opt_other}.

\begin{table}[h]
\centering
\caption{Optimal vs Fastest mode power consumption estimates on HL-LHC energies for all tested CPUs. Each number was calculated with 1000 configurations (5 million event each) and with 10 iterations and 4 collision energies.}
\label{tab:opt_other}
\begin{tabular*}{\textwidth}{@{\extracolsep{\fill}}l r r r}
\toprule
\textbf{CPU} & \textbf{Fastest (pp)} & \textbf{Optimal (pp)} & \textbf{Change} \\
\midrule
Intel Xeon E5-2650 & $2\, 990$ kWh  & $2\, 990$ kWh & 100\%  \\
AMD EPYC 7502P     & $2\, 514$ kWh  & $2\, 514$ kWh & 100\%  \\
AMD Ryzen 7 8845HS & $514$ kWh   & $476$ kWh  & 92.6\% \\
AMD EPYC 4585PX - SSD & $1\, 043$ kWh  & $972$ kWh & 93.2\% \\
AMD EPYC 4585PX - HDD & $1\, 111$ kWh  & $1\,085$ kWh & 97.7\% \\
AMD EPYC 4585PX - NW  & $314$ kWh & $314$ kWh & 100\% \\
\botrule
\end{tabular*}
\end{table}

The first clear observation is that in such a scenario the overall power requirement of the relatively older CPUs can be significantly higher than the newer ones. Additionally, on these more recent architectures there is a difference between the fastest and optimal parallelisation level, favoring the more optimal layouts by $\sim$7\%, which might lead to substantial energy saving on the long term. Finally, it might be surprising that the Ryzen 7 8845HS CPU used $\sim$50\% less energy than the more recent EPYC 4585PX in SSD and HDD mode---however, it is important to emphasize that the former is a laptop CPU designed for high level of efficiency, and consequently, a sustained level of such a heavy workload might reduce the lifespan of the hardware.

Here we only present the power consumption estimates for the proton-proton case---if one would run the MC event generator for heavy ion collisions, then the numbers would be inflated by several factors. Realistically speaking, at least $\sim 2$ million events would be required to tune the MC event generator for heavy-ion collisions, so the power consumption would be roughly $\mathcal{O}(100)$ times more.

\subsection{FCC era predictions}
\label{sec:tune2}

\begin{figure}[h]
  \centering
  \includegraphics[width=0.9\linewidth]{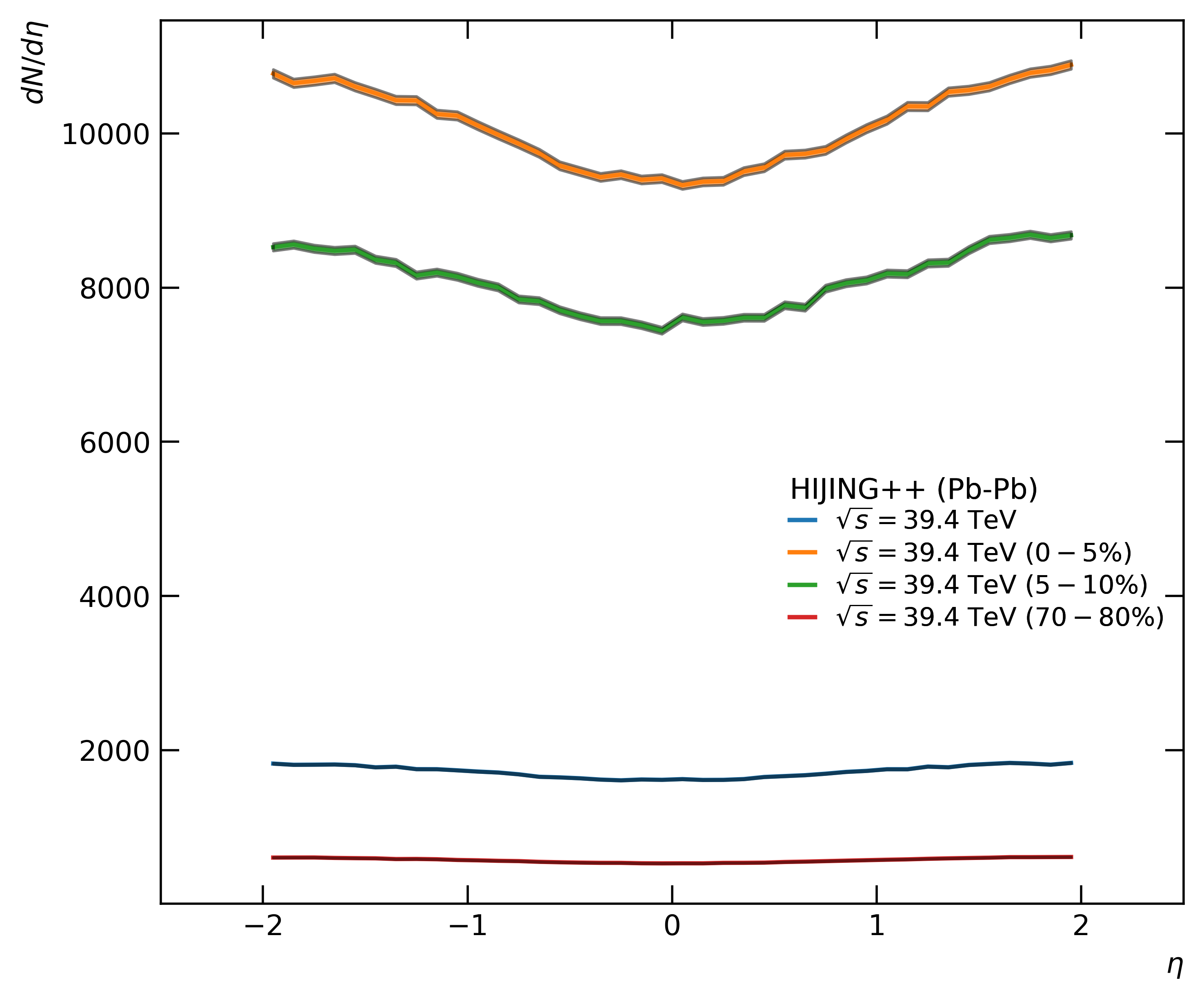}
  \caption{Pseudorapidity distribution predictions for 39.4 TeV Pb-Pb collisions with centrality classes (0-5\%, 5-10\%, 70-80\% and minimal bias).}
  \label{fig:tuneeta2}
\end{figure}

The \texttt{HIJING++} event generator code is still under development and under tuning for HL-LHC energies, but we can perform simulations for higher, FFC energies as well. Here the event generator predictions for $\sqrt{s_{NN}}=$39.4 TeV Pb-Pb collisions are presented: Fig.~\ref{fig:tuneeta2} shows the pseudorapidity distribution predictions of charged hadrons at various centrality classes.

The rough estimate for the power consumption of a future tuning scenario for proton-proton collisions at one FCC energy is shown  in Table~\ref{tab:opt_fcc}. Such a tuning process is expected to be more power hungry than for HL-LHC energies, but again, an impactful amount of energy can be conserved by optimizing the level of multithreading.

\begin{table}[h]
\centering
\caption{Optimal vs Fastest mode power consumption at FCC. Each number was calculated with 1000 configurations (5 million event each) with 10 iterations, at 1 energy value.}
\label{tab:opt_fcc}
\begin{tabular*}{\textwidth}{@{\extracolsep{\fill}}l r r r}
\toprule
\textbf{CPU} & \textbf{Fastest (pp)} & \textbf{Optimal (pp)} & \textbf{Change} \\
\midrule
AMD EPYC 4585PX - SSD & 531 kWh & 515 kWh & 97\% \\
AMD EPYC 4585PX - HDD & 506 kWh & 471 kWh & 93\% \\
AMD EPYC 4585PX - NW  & 176 kWh & 176 kWh & 100\% \\
\botrule
\end{tabular*}
\end{table}

\section{Discussion}
\label{sec:comp}

As it has been shown in Sec. \ref{sec:results1}, the optimal level of multithreading might be different than the fastest one, depending on the actual hardware. For the older CPUs, the optimal and the fastest cases can be the same. On the other hand, on the more recent AMD EPYC 4585PX CPU, the required cost of event generation can be reduced by 29\%, 33.65\% and 32.9\% for SSD, HDD and NW scenarios, respectively, by choosing the optimal configuration in Pb-Pb collision compared to the single core runs. However, the the actual runtime at the given level of multithreading is still to be investigated.

\begin{table}[h]
\caption{Multithread modes compared to the single thread mode. The measured runtimes of the single thread modes were taken as the reference (100\%).}
\label{tab:res_comp}
\begin{tabular*}{\textwidth}{@{\extracolsep{\fill}}l *{4}{c}}
\toprule
\multirow{2}{*}{\textbf{CPU}} &
  \multicolumn{2}{c}{\textbf{Change in time - fastest}} &
  \multicolumn{2}{c}{\textbf{Change in time - optimal}} \\
& pp & Pb-Pb & pp & Pb-Pb \\
\midrule
Intel Xeon E5-2650       & $67.2\%$ & $59.3\%$ & $67.2\%$ & $59.3\%$ \\
AMD EPYC 7502P           & $75.4\%$ & $59.0\%$ & $75.4\%$ & $59.0\%$ \\
AMD Ryzen 7 8845HS       & $49.8\%$ & $49.9\%$ & $51.8\%$ & $56.8\%$ \\
AMD EPYC 4585PX - SSD    & $44.0\%$ & $81.1\%$ & $50.3\%$ & $81.1\%$ \\
AMD EPYC 4585PX - HDD    & $43.6\%$ & $92.4\%$ & $47.3\%$ & $92.4\%$ \\
AMD EPYC 4585PX - NW     & $19.7\%$ & $93.9\%$ & $19.7\%$ & $93.9\%$ \\
\bottomrule
\end{tabular*}
\end{table}

Table~\ref{tab:res_comp} compares the runtime of the \texttt{HIJING++} event generator in the two multithreaded configurations with respect to the single thread case. As expected, the multithreaded mode (either in the fastest or the optimal configuration) results in a substantial speedup in every case. What is more important, this speedup is very similar between the fastest and optimal cases (except for the older CPUs, where these two coincide), the runtimes of the optimal runs lag only a few percents behind. 
The amount of the saved energy can be notable in the optimal case, leading to prominent energy cost reduction---the tradeoff is only a negligible increase in runtime.

\begin{table}[h]
\caption{Multithread modes compared to the single thread mode, FCC energies. The measured runtimes of the single thread modes were taken as the reference (100\%).}
\label{tab:res_comp_4585px_FCC}
\begin{tabular*}{\textwidth}{@{\extracolsep{\fill}}l *{4}{c}}
\toprule
\multirow{2}{*}{\textbf{CPU - Drive}} & 
\multicolumn{2}{c}{\textbf{Change in time - fastest}} &
\multicolumn{2}{c}{\textbf{Change in time - optimal}} \\
& pp & Pb-Pb & pp & Pb-Pb \\
\midrule
AMD EPYC 4585PX - SSD    & $43.98\%$ & $81.11\%$ & $50.26\%$ & $81.11\%$ \\
AMD EPYC 4585PX - HDD    & $43.59\%$ & $92.39\%$ & $47.33\%$ & $92.39\%$ \\
AMD EPYC 4585PX - NW     & $19.66\%$ & $93.94\%$ & $19.66\%$ & $93.94\%$ \\
\bottomrule
\end{tabular*}
\end{table}

As it has been shown on Fig.~\ref{fig:4585px_FCC}, the energy consumption, runtime and multithreading relations are similar even at FCC collision energies. In Table~\ref{tab:res_comp_4585px_FCC} we show the same runtime comparisons with respect to the single threaded mode for the EPYC 4585PX CPU. As it was in the HL-LHC era, a good amount of speedup can be achieved with the multithreading---however, this speedup is more moderate for the Pb-Pb collisions.

In all of the above detailed cases we utilized the \texttt{HIJING++} Monte Carlo event generator with a demonstrative tuning scenario to show that the level of multithreading can be optimized compared to the naive performance gain, therefore the energy footprint can be reduced. We investigated a handful of various CPU architectures---however, the same principle can be generalized and applied to other parallelizable calculations and hardware configurations as well. Eventually, this approach might ease the forthcoming computational challenges and lead to a more sustainable operation of event generators for the HL-LHC and FCC era.

\section{Summary}
\label{sec:sum}
Given the recent increase in computational requirements worldwide, the question of how multithreading impacts the sustainability and efficiency of Monte Carlo event generators led to the conclusions of this study. Various CPUs were tested to determine the optimal and most efficient level of parallelisation.

Using the Toolbox, which includes all the necessary tools for the development and tuning of Monte Carlo event generators like \texttt{HIJING++}, can not only speed up the process but also provide a feasible method for comparing hardware. The \mbox{77rev/proripy:4.4} Docker image contains also the benchmarking script(s) that made it possible.

We have shown that the given CPU architecture and the optimal level of multithreading play a significant role in the sustainable operation of MC event generators. The newer hardware is proven to be more efficient: we tested the AMD EPYC 4585PX the most extensively. The difference between drive type (SSD vs HDD) revealed only slight differences.

Additionally, the AMD EPYC 4585PX has been tested and benchmarked for FCC energies at $\sqrt{s}=$50 TeV proton-proton, and $\sqrt{s_{NN}}=$39.4 TeV Pb-Pb collisions, leading to similar conclusions: for the beyond LHC era of collider experiments, the efficient usage of the available computational resources plays a crucial role in sustainability.

\section*{Acknowledgement}

The authors would like to thank the support of the Hungarian National Research, Development and Innovation Office (NKFIH) grants under the contract numbers NKKP ADVANCED\_25-153456, 2025-1.1.5-NEMZ\_KI-2025-00005, 2025-1.1.5-NEMZ\_KI-2025-00013, 2024-1.2.5-TÉT-2024-00022. Computational resources were provided by the  Wigner Scientific Computing Laboratory (WSCLAB). 

\bibliography{references}

\begin{appendices}
\label{sec:app} 

\section{Additional plots and tables}
\label{sec:app-plottab}

\begin{figure}[h]
  \centering
  \includegraphics[width=1\columnwidth]{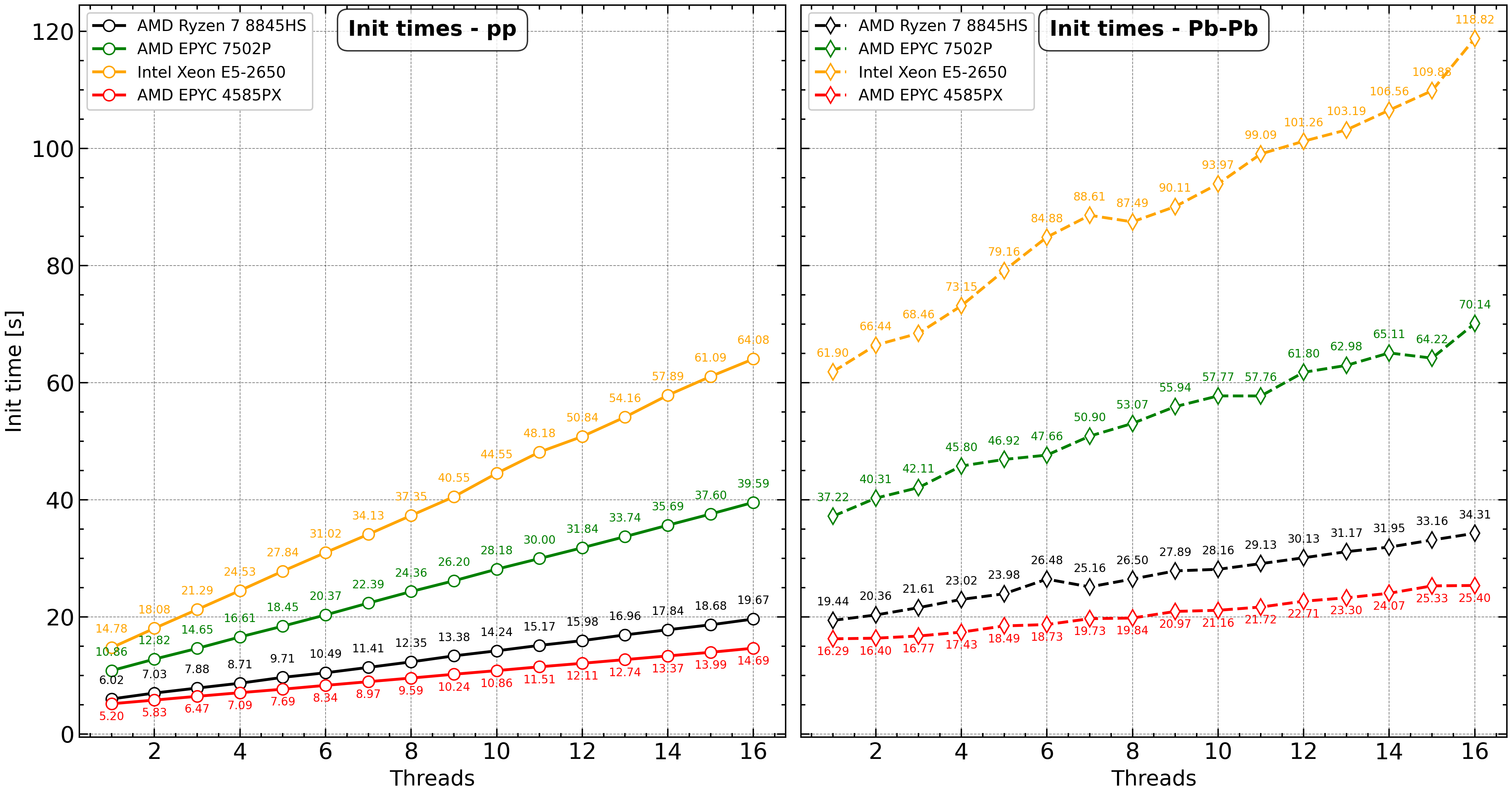}
  \caption{Initialization times per CPU, from Table~\ref{tab:inits}}
  \label{fig:inits}
\end{figure}

Initialization is required to set up the calculations and the objects that the code uses. 
Table~\ref{tab:inits} shows these times in a table format, while Fig.~\ref{fig:inits} shows it in a plot format.

\begin{table}[h]
\caption{Initialization times across CPUs per thread and per pp/Pb-Pb collisions}\label{tab:inits}
\begin{tabular*}{\textwidth}{@{\extracolsep\fill} l l *{8}{r} }
\toprule
\multirow{2}{*}{CPU} & \multirow{2}{*}{Thread} & 1 & 2 & 3 & 4 & 5 & 6 & 7 & 8 \\
    & & 9 & 10 & 11 & 12 & 13 & 14 & 15 & 16 \\
\midrule
\multirow{4}{*}{Xeon E5-2650} & \multirow{2}{*}{pp [s]} & 14.78 & 18.08 & 21.29 & 24.53 & 27.84 & 31.02 & 34.13 & 37.35 \\
                              &                      & 40.55 & 44.55 & 48.18 & 50.84 & 54.16 & 57.89 & 61.09 & 64.08 \\
                              & \multirow{2}{*}{Pb-Pb [s]} & 61.9  & 66.44 & 68.46 & 73.15 & 79.16 & 84.88 & 88.61 & 87.49 \\
                              &                      & 90.11 & 93.97 & 99.09 & 101.26 & 103.19 & 106.56 & 109.88 & 118.82 \\
\midrule
\multirow{4}{*}{EPYC 7502P} & \multirow{2}{*}{pp [s]} & 10.86 & 12.82 & 14.65 & 16.61 & 18.45 & 20.37 & 22.39 & 24.36 \\
                            &                      & 26.2  & 28.18 & 30    & 31.84 & 33.74 & 35.69 & 37.6  & 39.59 \\
                            & \multirow{2}{*}{Pb-Pb [s]} & 37.22 & 40.31 & 42.11 & 45.8  & 46.92 & 47.66 & 50.9  & 53.07 \\
                            &                      & 55.94 & 57.77 & 57.76 & 61.8  & 62.98 & 65.11 & 64.22 & 70.14 \\
\midrule
\multirow{4}{*}{Ryzen 7 8845HS} & \multirow{2}{*}{pp [s]} & 6.02  & 7.03  & 7.88  & 8.71  & 9.71  & 10.49 & 11.41 & 12.35 \\
                                &                      & 13.38 & 14.24 & 15.17 & 15.98 & 16.96 & 17.84 & 18.68 & 19.67 \\
                                & \multirow{2}{*}{Pb-Pb [s]} & 19.44 & 20.36 & 21.61 & 23.02 & 23.98 & 26.48 & 25.16 & 26.5  \\
                                &                      & 27.89 & 28.16 & 29.13 & 30.13 & 31.17 & 31.95 & 33.16 & 34.31 \\
\midrule
\multirow{4}{*}{EPYC 4585PX} & \multirow{2}{*}{pp [s]}  & 5.2 & 5.83 & 6.47 & 7.09 & 7.69 & 8.34 & 8.97 & 9.59 \\
                             &                      & 10.24 & 10.86 & 11.51 & 12.11 & 12.74 & 13.37 & 13.99 & 14.69 \\
                             & \multirow{2}{*}{Pb-Pb [s]}  & 16.29 & 16.4 & 16.77 & 17.43 & 18.49 & 18.73 & 19.73 & 19.84 \\
                             &                      & 20.97 & 21.16 & 21.72 & 22.71 & 23.3 & 24.07 & 25.33 & 25.4 \\
\bottomrule
\end{tabular*}
\end{table}

\begin{figure}[h]
  \centering
  \includegraphics[width=1\columnwidth]{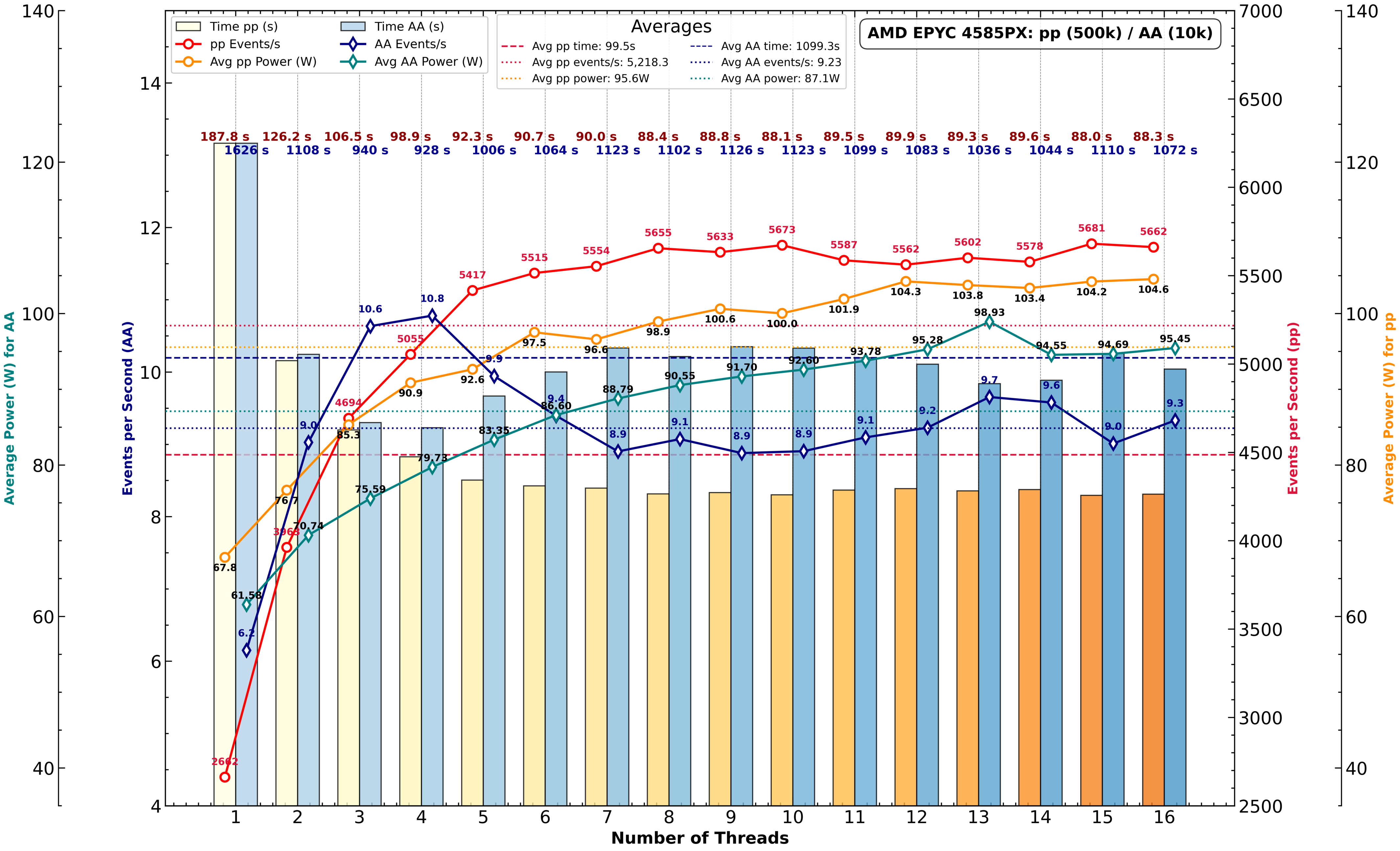}
  \caption{Detailed results for the AMD EPYC 4585PX on SSD}
  \label{fig:detEPYC4585PX}
\end{figure}

\begin{figure}[h]
  \centering
  \includegraphics[width=1\columnwidth]{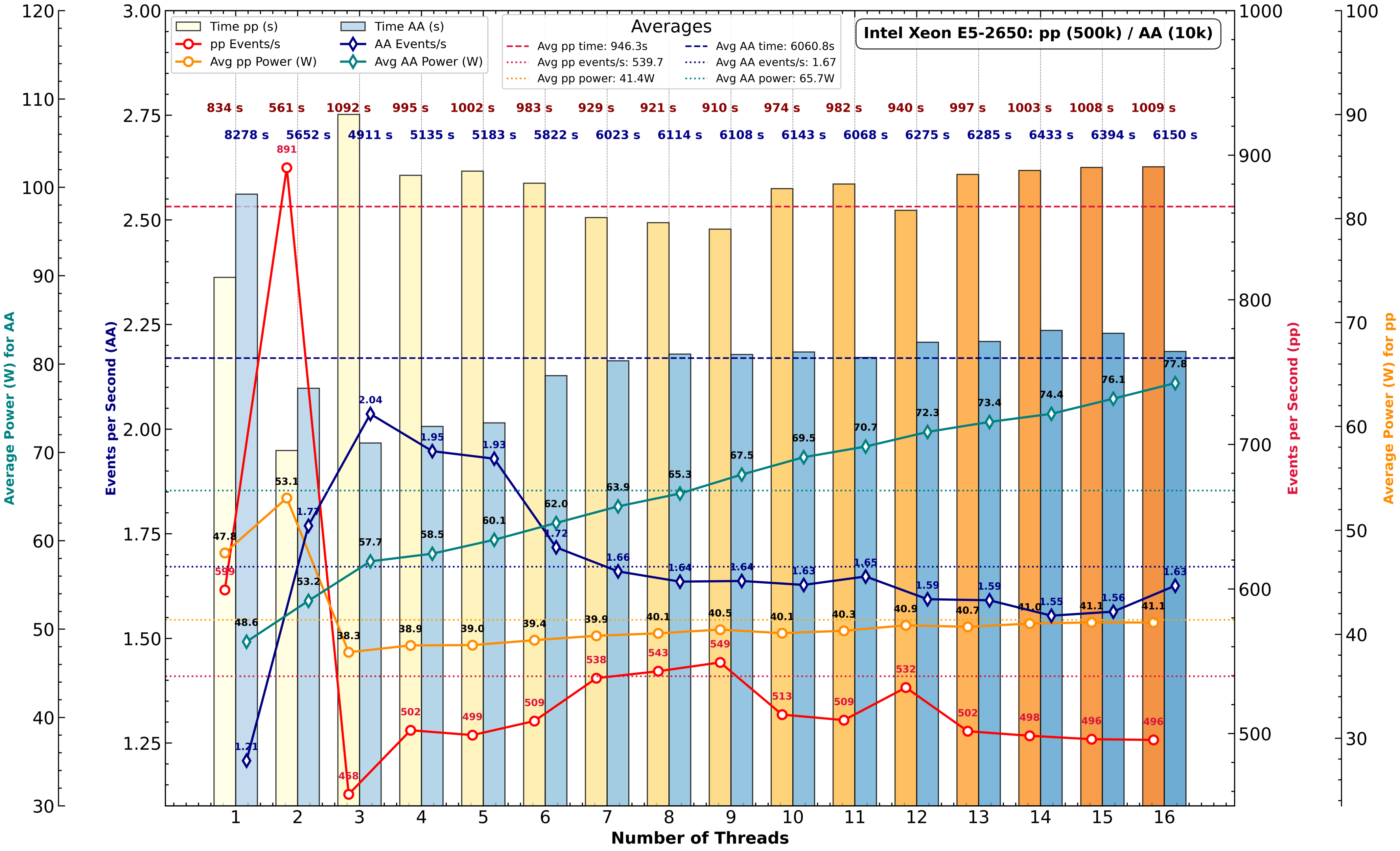}
  \caption{Performance metrics of the Intel Xenon E5-2650, on HDD}
  \label{fig:xeon}
\end{figure}

\begin{figure}[h]
  \centering
  \includegraphics[width=1\columnwidth]{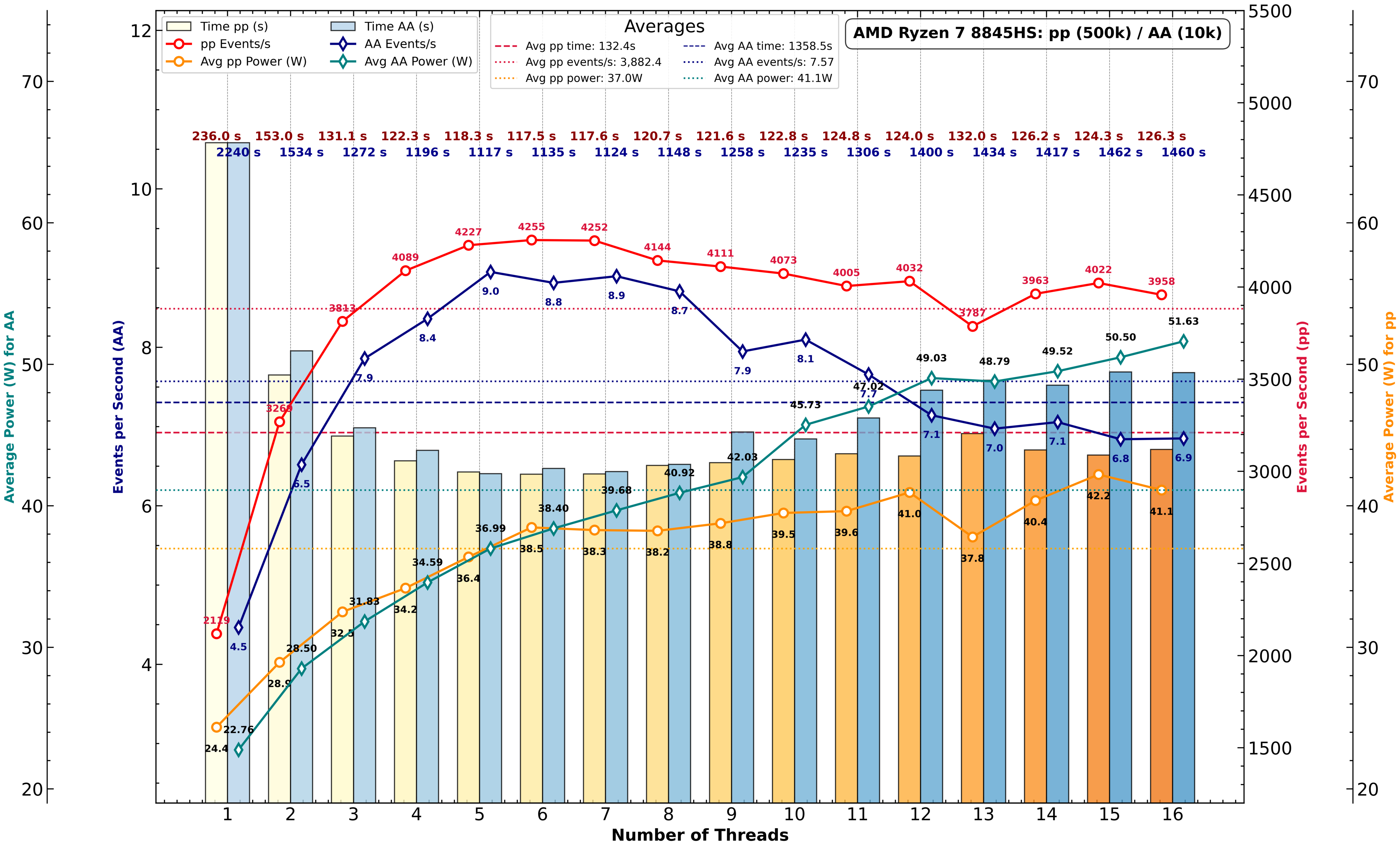}
  \caption{Performance metrics of the AMD Ryzen 7 8845HS, on SSD}
  \label{fig:ryzen}
\end{figure}

\begin{figure}[h]
  \centering
  \includegraphics[width=1\columnwidth]{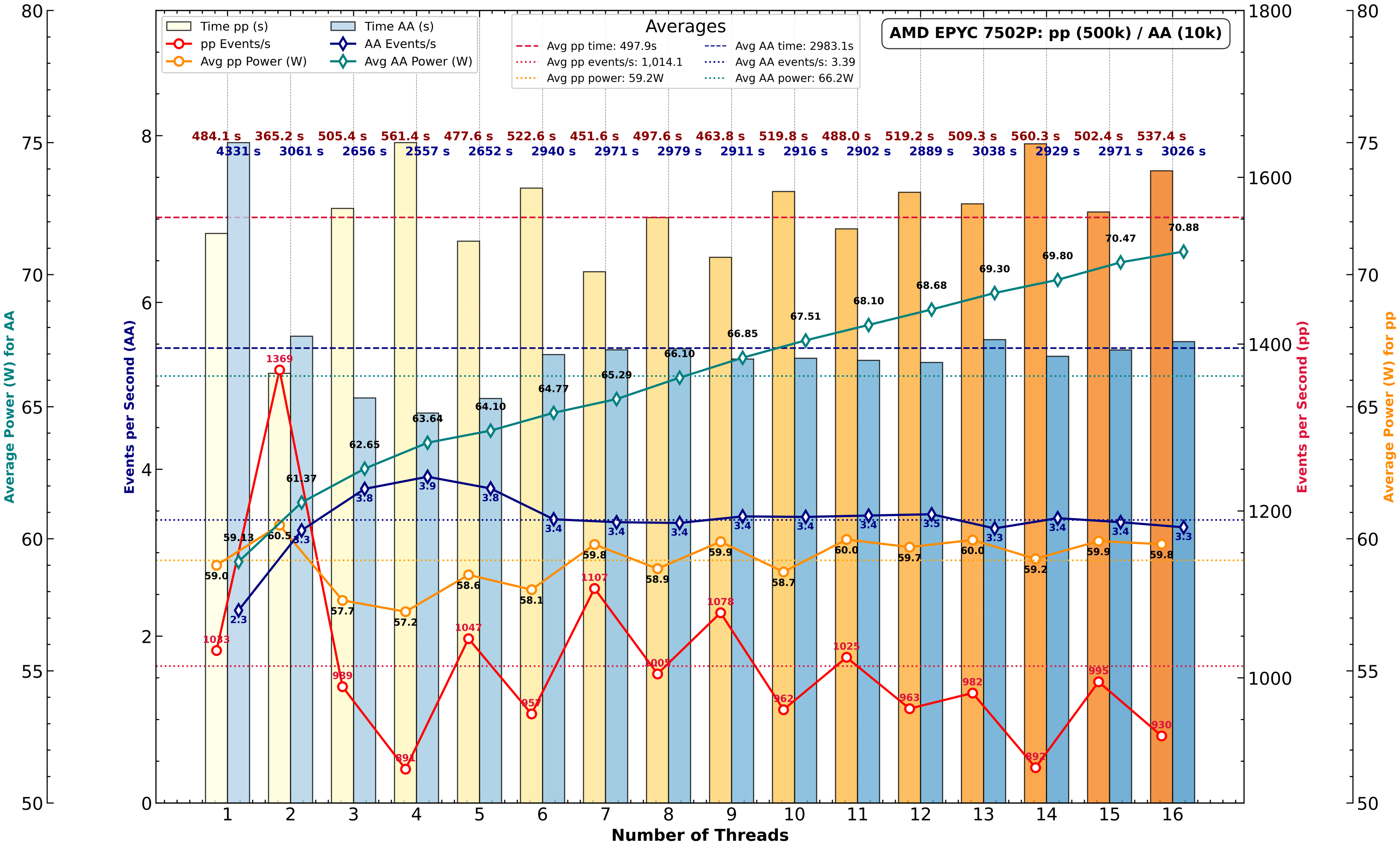}
  \caption{Performance metrics of the AMD EPYC 7502P, on HDD}
  \label{fig:7502p}
\end{figure}

We also show the detailed benchmarking result breakdowns in Fig~\ref{fig:detEPYC4585PX}-\ref{fig:7502p}.

\end{appendices}
\end{document}